\documentstyle[12pt]{article}

\begin{document}

\vspace*{3cm}

\begin{center}
{\Large {\bf Equivariant Localization: BV-geometry and Supersymmetric
Dynamics }}\\ \bigskip
\bigskip
\bigskip
\bigskip
{\large {\bf Armen Nersessian}}{\
\footnote
{On leave of absence from University of Karabakh, Stepanakert,
Nagorny Karabakh (Former Soviet
Union)}$^,$\footnote{E-MAIL:NERSESS@THEOR.JINRC.DUBNA.SU}} \bigskip
\bigskip
\vspace*{0.8cm}\\ {\it Laboratory of Theoretical Physics, JINR}\\ {\it %
Dubna, Head Post Office, P.O.Box 79, 101 000 Moscow, Russia}\\
\end{center}

\bigskip
\bigskip
\bigskip
\bigskip
\begin{abstract}
It is shown, that the geometrical objects of Batalin-Vilkovisky formalism--
odd symplectic structure and nilpotent operator $\Delta$ can be naturally
uncorporated in Duistermaat--Heckman localization procedure. The presence of
the supersymmetric bi-Hamiltonian dynamics with even and odd symplectic
structure in this procedure is established. These constructions can be
straightly generalized for the path-integral case.
\end{abstract}

\vfill


\setcounter{page}0
\renewcommand{\thefootnote}{\arabic{footnote}} \setcounter{footnote}0
\newpage
\setcounter{equation}0

\section{Introduction}

Recently, in a number of papers ( cf. \cite{niemi1}--\cite{witten}) the exact
evaluation of the phase space path integrals was studied.
The (formal) path integral generalization \cite{niemi1} of the
Duistermaat-Heckman localization formula (DH-formula) \cite{DH}, \cite
{atbott} was used for this purpose . Due to this formula if $M$ is the
compact $2N$-dimensional symplectic manifold with the symplectic structure $%
\omega =\frac 1{2\pi }\omega _{ij}dx^i\wedge dx^j$ and Hamiltonian $H$
define the circle action on $M$ then
\begin{equation}
\label{DH}\int_M{\rm e}^H(\omega )^N=\sum_{dH=0}\frac{{\rm e}^H\sqrt{
det\omega _{ij}}}{\sqrt{det\frac{\partial ^2H}{\partial x^i\partial x^j}}}.
\end{equation}

This formula was applied at first to the evaluation of path integral in
Ref.\cite {atiah} where Dirac's operator index was calculated.

This approach turns out to be convenient for a large class of problems \cite
{niemi2}-\cite{witten} including conceptually new, geometric interpretation
of supersymmetric theories \cite{morozov} and revision of two-dimensional
Yang-Mills theory \cite{witten}.

The derivation of (\ref{DH}) can be formulated in terms of supersymmetry
which corresponds to the {\it equivariant} {\it transformations} \cite
{atbott}. Namely, the r. h. s. of (\ref{DH}) can be presented in the form
\begin{equation}
\label{int}Z_0=\frac 1{(2\pi )^N}\int_M{\rm e}^{H(x)}\sqrt{det\omega _{ij}}
d^{2N}x=\frac 1{(\pi )^N}\int_{{\cal M}}\exp ({H(x)+\frac 12\omega
_{ij}\theta ^i\theta ^j)}d^{2N}xd^{2N}\theta ,
\end{equation}
where $\theta ^i$ are auxiliary Grassmannian fields corresponded to the
basic 1-forms $dx^i$, ${\cal M}$ is the supermanifold associated with the
tangent bundle of $M$.
With respect to $\theta ^i\leftrightarrow dx^i$, to
any differential form $\omega ^k=\omega _{i_1...i_k}^kdx^{i_1}\wedge
...\wedge dx^{i_k}$ on $M$ the monomial ${\tilde \omega }^k=\omega
_{i_1...i_k}^k\theta ^{i_1}...\theta ^{i_k}$ on ${\cal M\ }$is corresponded.

Since $H$ defines on $M$ a circle action one can construct the metric $%
g_{ij} $ and the odd function (''gauge fermion'')
\begin{equation}
\label{eq:gauge}{\tilde Q}_H=\xi _H^ig_{ij}\theta ^j,
\end{equation}
which are both Lie-invariant with respect to $\xi _H^i$, where
\begin{equation}
\label{xi}\xi _H^i\equiv \frac{\partial H}{\partial x^j}\omega ^{ji}.
\end{equation}
Then, both integrals: (\ref{int}) and
\begin{equation}
\label{eq:int2}Z_\lambda =\int_{{\cal M}}\exp (H+\frac 12\omega _{ij}\theta
^i\theta ^j-\lambda \hat d_H{\tilde Q}_H)d^{2N}d^{2N}\theta ,
\end{equation}
(where $\hat d_H=\hat d+\hat \imath_H$ is equivariant differential along $\xi
_H^i $, and $\lambda $ is a numerical parameter) are invariant under
supersymmetry transformations generated by the equivariant differential $%
\hat d_H=\hat d+\hat \imath_H$ :
\begin{equation}
\label{eq:susy}\delta x^i=\theta ^i,\quad \delta \theta ^i=\xi _H^i.
\end{equation}
Moreover, the integral (\ref{eq:int2}) is $\lambda $ --independent.

Taking the limits $\lambda \to 0$ and $\lambda \to \infty $ we obtain the
standard DH-formula (\ref{DH}).

By this reason, further, we will call by DH-supermanifold the one associated
with tangent bundle of a compact manifold which is provided by Hamiltonian
dynamics and by Riemannian structure, Lie-invariant with respect to this
dynamics .

We will denote by ${\cal M}$ a supermanifold associated with the tangent
bundle of symplectic manifold $M$.

Since any function on ${\cal M}$ can be interpreted in terms of differential
forms on $M$ one can formulate the exterior differential calculus on $M$ in
terms of functions on ${\cal M}$.

The derivation of DH-formula (\ref{DH}) presented above was used in Ref.\cite
{niemi1}--\cite{niemi3} for the path integral generalization of DH-formula.
For this purpose, the formalism developed by E. Gozzi et al. for the path
integral formulation of classical mechanics \cite{gozzi} is used .
In this formalism the canonical symplectic structure constructed on the
cotangent bundle of ${\cal M}$ is used.

However, there is the alternative possibility of Hamiltonian description of
DH-lo\-ca\-li\-za\-tion which allows to avoid the introduction of the
structure of cotangent bundle of ${\cal M}$ by using the {\it odd symplectic
structure} constructed on ${\cal M}$ \cite{jetp}.

In this work we will present this approach more completely.

We will show that ${\cal M}$ can be provided with the basic objects of
BV-formalism \cite{bat}, \cite{bat2}: the odd symplectic structure
$\Omega_1$ and the nilpotent operator $\Delta $.

Using the symplectic structure $\omega $ on $M$ we define the natural lift
of an arbitrary Hamiltonian mechanics $(H,\omega ,M)$ on the certain odd one
$(Q_H, \Omega _1, {\cal M})$ which is defined by the odd function $Q_H$ and
the odd symplectic structure $\Omega _1$ on ${\cal M}$. This odd mechanics
defines the Lie derivatives of differential forms along $\xi _H^i$.
Moreover, it is invariant under supersymmetry transformations (\ref{eq:susy}%
) which are generated on the odd symplectic structure by the initial
Hamiltonian $H$ and by the symplectic structure $\omega $. If ${\cal M}$ is
the DH-supermanifold then the ''gauge fermion'' (\ref{eq:gauge}) plays the
role of the additional (odd) motion integral of $(Q_H,\Omega _1,{\cal M})$.

Hence this odd Hamiltonian mechanics gives the natural description of the
symmetries of the integrals (\ref{int}), (\ref{eq:int2}) and, therefore, of
the DH-localization.

The existence of Riemannian metric on $M$ allow us to construct on
${\cal M}$ the {\it even symplectic structure}. We show that if ${\cal M}$
is DH-supermanifold then one can construct the second, {\it even Hamiltonian
structure} for $(Q_H,\Omega _1,{\cal M})$.

It is interesting to point out that such bi-Hamiltonian systems (with even
and odd symplectic structures) were studied earlier without any connection
with DH-formula. The example of such a system (one-dimensional
supersymmetric Witten's mechanics) was present at first by D. V. Volkov et
al. \cite{volkov}. Later such systems were studied in more details in
Ref.\cite{pr}--\cite{red}. Particularly, in Ref. \cite{pr} such structure was
studied for the supersymmetric integrable mechanics in terms of
''action-angle" variables. In Ref.\cite{khud} there is the proof, that only
a finite number of such bi-Hamiltonian systems for fixed even and odd
symplectic structures exist. The bi-Hamiltonian systems with even and odd
symplectic structures on the superspaces were studied in Ref.\cite{jmp}, and
on K\"ahlerian supermanifolds in Ref.\cite{tmp} . In Ref.\cite{red} the
procedure of Hamiltonian reduction of such systems to complex projective
superspaces was considered.

Let us cite also the Ref.\cite{km} where it is shown that odd symplectic
structure has nontrivial geometrical properties with no analogues in the
even symplectic structure .

But the odd symplectic structure (with the connected object-- operator $%
\Delta $) is well-known in physics as the basic object of the covariant
quantization formalism for the theories with arbitrary constraints--
Batalin-Vilkovisky formalism (BV-formalism) \cite{bat}, \cite{bat2}.
 BV-formalism is considered now as the background in the construction of
invariant
string field theory (cf.\cite{wit2}). This is the stimulative factor in the
investigation of the geometry of BV-formalism and the odd symplectic
structure \cite{schwarz}.

However, the odd symplectic structure is used in BV--formalism for the
formulation of so-called ''master equation'' which is {\it kinematic} one.

As we noted above, the equivariant localization gives the example of
applying of the odd symplectic structure for the formulation of {\it %
dynamical} (Hamiltonian) equations (the existence of Hamiltonian systems
with odd symplectic structure was first pointed out in \cite{leites}).

The paper is organized as follows.

In {\it Section 2} we present the basic properties of the odd symplectic
structure and operator $\Delta$ and construct theirs on the supermanifold $%
{\cal M}$ associated with the tangent bundle of the symplectic manifold $M$ .

In {\it Section 3} we map (lift) the arbitrary Hamiltonian mechanics $(H,
\omega , M)$ on the odd supersymmetric Hamiltonian mechanics $(Q_H,
\Omega_1, {\cal M})$ and interpret its supersymmetry in terms of equivariant
transformations. We demonstrate that if ${\cal M}$ is DH-supermanifold then
one can define the even Hamiltonian structure on it determining the same
dynamics as $(Q_H, \Omega _1, {\cal M}).$

In {\it Section 4} using the results of the previous sections we repeat the
derivation of DH-localization formula.

For rigorous definitions and conventions in
on Hamiltonian mechanics and symplectic geometry used here we refer to
Ref.\cite{arnold}, and Ref.\cite{ber} on supergeometry and superanalyses .

\setcounter{equation}0

\section{BV-Geometry}

Odd Poisson bracket ( antibracket) of functions $f(z)$ and $g(z)$ on the
supermanifold is the bi-linear differential operation
\begin{equation}
\label{eq:bloc}\{f,g\}_1=\frac{\partial _rf}{\partial z^A}\Omega _1^{AB}(x)
\frac{\partial _lg}{\partial z^B}\,\,\,\,\,\,,
\end{equation}
which satisfy the conditions
\begin{eqnarray}
& &p(\{ f, g \}_{1} )= p(f)+ p(g) + 1   \quad {\rm (grading \quad condition)} ,
\nonumber \\
& &\{ f, g \}_1 = -(-1)^{(p(f)+1)(p(g)+1)}\{ g, f \}_1
\quad {\rm ( "antisymmetricity")} ,\label{eq:anti} \\
& & ( -1)^{(p(f)+1)(p(h)+1)}\{ f,\{ g, h \}_{1}\}_1 +
{\rm {cycl. perm. (f, g, h)}} = 0
  \quad{\rm {(Jacobi\quad id.)}} ,
\label{eq:bjac} \end{eqnarray}
where $z^A$ are the local coordinates , and
$\frac{\partial _r}{\partial z^A} $ and $\frac{\partial _l}{\partial z^A}$
denotes correspondingly right and left derivatives.

On the supermanifold with an equal number of even and odd coordinates
 the odd bracket can be nondegenerate one. Then one can corresponds to it the
odd
symplectic structure
\begin{equation}
\label{eq:symp}\Omega _1=dz^A\Omega _{(1)AB}dz^B
\end{equation}
where $\Omega _{(1)AB}\Omega _1^{BC}=\delta _A^C$. Locally, the
nondegenerate odd Poisson bracket can be transformed to the canonical form
\cite{leites}:
\begin{equation}
\label{eq:bcan}\{f,g\}_1^{{\rm can}}=\sum_{i=1}^N\left( \frac{\partial _rf}{%
\partial x^i}\frac{\partial _lg}{\partial \theta _i}-\frac{\partial _rf}{%
\partial \theta _i}\frac{\partial _lg}{\partial x^i}\right) ,
\end{equation}
where $p(\theta _i)=p(x^i)+1$.

It was shown in Ref.\cite{bat}, \cite{km} that the odd Poisson bracket has not
invariant volume forms. Hence, if the supermanifold is provided with the
odd bracket (\ref{eq:bloc}) and with the volume form
\begin{equation}
\label{eq:vol}dv=e^{\rho (z)}d^{2N}z\,\,\,\,\,\,\,,
\end{equation}
where $e^\rho $-- some integral density, then one can define on
it the odd differential operator of the second order, so called ``operator $%
\Delta $'' which is invariant under the transformations conserving the
symplectic structure and the volume form \cite{khud}. Its action on the
function $f(x,\theta )$ is the divergency of the Hamiltonian vector field $%
{\bf D}_f=\{\quad ,f\}_1$ with the volume form (\ref{eq:vol}):
\begin{equation}
\label{eq:delta}\Delta _\rho f=\frac 12div_\rho {\bf D}_f\equiv \frac 12
\frac{{\cal L}_{{\bf D}_f}dv}{dv},
\end{equation}
where ${\cal L}_{{\bf D}_f}$ --- Lie derivative along ${\bf D}_f$ , or in
the coordinate form
\begin{equation}
\label{eq:deltaloc}\Delta _\rho f=\frac 12\frac{\partial ^R}{\partial z^A}%
\left( \{z^A,f\}_1\right) +\frac 12\{\rho ,f\}_1\equiv \Delta _0+\frac
12\{\rho ,f\}_1   .
\end{equation}

The operator $\Delta $ which is used in Batalin-Vilkovisky formalism obeys
to the nilpotency condition
\begin{equation}
\Delta _\rho ^2=0,
\end{equation}
which holds if the density $\rho (z)$ satisfy some conditions. In more
details the properties of $\Delta $ are considered in Ref. \cite{schwarz}.

It is known that any supermanifold one can associate with some vector bundle
\cite{ber}. On the supermanifold associated with the cotangent bundle of any
manifold it can be constructed the odd symplectic structure \cite{leites}.

Indeed, let $T^{*}M$ be the cotangent bundle to the manifold $M$, $x^i$ are
the local coordinates on $M$ and $(x^i,\theta _i)$ are the corresponding
local coordinates on $T^{*}M$ with the transition functions
\begin{equation}
\label{eq:trans} {\tilde x}^i={\tilde x}^i(x),\quad
{\tilde \theta }_i=\sum_{i=1}^N\frac{\partial x^j}{\partial {%
\tilde x}^i}\theta _j.
\end{equation}
Considering for every map the superalgebra generated by $(x^i,\theta _i)$
where $x^i$ are even and $\theta _i$ are odd, transforming from map to map
like $(x^i,\theta _i)$ in the (\ref{eq:trans}) we go to supermanifold ${\cal %
M}$ which is associated to $T^{*}M$ in the coordinates $(x^i,\theta _i) $ .
Obviously, on this supermanifold in coordinates $(x^i,\theta _i)$ one can
globally define the canonical odd symplectic structure with the canonical
odd bracket (\ref{eq:bcan}) . (By the same way one can define an odd Poisson
bracket on the supermanifold associated with cotangent bundle of a {\it %
supermanifold}.)

Let us construct an odd symplectic structure and the operator $\Delta $ on
the supermanifold ${\cal M}$ associated with the tangent bundle of the
symplectic manifold $M$.

Let
\begin{equation}
\label{eq:beven}\{f(x),g(x)\}=\frac{\partial f}{\partial x^i}\omega ^{ij}
\frac{\partial g}{\partial x^j}
\end{equation}
is the nondegenerate Poisson bracket on $M$.

Then the function
\begin{equation}
\label{eq:F0}F(z)=\frac 12\theta _i\omega ^{ij}\theta _j=-\frac 12\theta
^i\omega _{ij}\theta ^j
\end{equation}
is globally defined on ${\cal M}$ and satisfy the condition
\begin{equation}
\label{eq:ff}\{F,F\}_1=0.
\end{equation}
 Using this function one can map any
function $f(x)$ on $M$ to ${\cal M}$:
\begin{equation}
\label{eq:map}f(x)\to Q_f(z)=\{f(x),F(x,\theta )\}_1\,\,\,\,\,\,.
\end{equation}
This map has the following important property
\begin{equation}
\label{eq:cons}\{f(x),g(x)\}=\{f(x),Q_g(x,\theta )\}_1\quad {\rm for}\quad
{\rm any}\quad f(x),g(x) .
\end{equation}
Then doing the coordinate transformation
\begin{equation}
\label{eq:theta}(x^i,\theta _i)\to (x^i,\theta ^i=\{x^i,F(z)\}_1)\,\,,
\end{equation}
 From (\ref{eq:theta}) we see that $\theta ^i$ transforms like $dx^i$, then
it can be interpreted as the basis of 1-form on $M$. Hence, $z^A=(x^i,\theta
^i)$ plays the role of the local coordinates of the supermanifold associated
with tangent bundle of the symplectic manifold $M$.

An odd symplectic structure in these coordinates takes the form
\begin{equation}
\label{eq:osym}\Omega_{1} =\omega_{ij}dx^i\wedge d\theta^j
+\omega_{ij,k}\theta^j dx^k\wedge dx^j.
\end{equation}

The corresponding odd Poisson bracket is defined by following basic
relations:

\begin{equation}
\label{eq:bxt}\{x^i,x^j\}_1=0,\quad \{x^i,\theta ^j\}_1=\omega ^{ij},\quad
\{\theta ^i,\theta ^j\}_1=\frac{\partial \omega ^{ij}}{\partial x^k}\theta
^k,
\end{equation}
where $\omega ^{ij}$ is the matrix of the even Poisson bracket (\ref
{eq:beven}) on $M$. The operator $\Delta $ connected with it takes the form
\begin{equation}
\label{delta1}\Delta _\rho =\omega ^{ij}\frac{\partial ^2}{\partial
x^i\partial \theta^j}+\frac{1}{2}\omega _{,k}^{ij}\theta ^k\frac{\partial
^2}{\partial
\theta ^i\partial \theta ^j}+\frac 12\{\rho ,\quad \}_1 .
\end{equation}
It is easy to check that if
\begin{equation}
\rho =kF
\end{equation}
( $k$ is an arbitrary numerical constant) then this operator is nilpotent
one:
\begin{equation}
\Delta _{kF}^2=0.
\end{equation}
Thus, we provide the supermanifold ${\cal M}$ associated with tangent bundle
of symplectic manifold $M$ with the geometrical structures of
Batalin-Vilkovisky formalism. As we noticed in Introduction, such
supermanifold presents in the Duistermaat-Heckman localization procedure.

In the next Section we will study the lift (\ref{eq:map}) of the Hamiltonian
mechanics on the symplectic manifold.

\setcounter{equation}0

\section{Bi-Hamiltonian Dynamics}
Let the symplectic manifold $M$ is provided with the Hamiltonian mechanics $%
\left( H(x),\omega ,M\right) ,$ where the symplectic structure $\omega $
defines the nondegenerate Poisson bracket (\ref{eq:beven}) on $M$ and $H(x)$
is the Hamiltonian on it.

Using (\ref{eq:map}), (\ref{eq:cons}) let us map this mechanics on $\left(
Q_H,\Omega _1,{\cal M}\right) $, where
\begin{equation}
\label{eq:q}Q_H=\{H,F\}_1
\end{equation}
is the odd Hamiltonian on ${\cal M}$ (the mapping of Hamiltonian $H$ on $%
{\cal M}$), and $\Omega _1$ is the odd symplectic structure (\ref{eq:osym})
on it which defines the antibracket (\ref{eq:bxt}).

It is easy to see that the equations of motion of $\left( Q_H,\Omega _1,%
{\cal M}\right) $ in terms of $(x^i,\theta ^i)$ takes the form:
\begin{equation}
\label{eq:motion}\frac{dx^i}{dt}=\{x^i,Q_H\}_1=\xi _H^i,\quad\quad\quad
\\\frac{%
d\theta ^i}{dt}=\{\theta ^i,Q_H\}_1=\frac{\partial \xi _H^i}{\partial x^j}%
\theta ^j.
\end{equation}

This mechanics is supersymmetric one . Indeed, using (\ref{eq:ff}) (\ref
{eq:q}) and definition of $H$ we see that $H,F,Q_H$ forms the simplest
superalgebra:
\begin{eqnarray}
 && \{H\pm F, H \pm F \}_{1} =\pm 2Q_{H} ,\\
 && \{H+ F, H -  F\}_{1} = \{H\pm F, Q_{H} \}_{1}
= \{Q_{H}, Q_{H} \}_{1} = 0                           \nonumber
\label{eq:sualg2} \end{eqnarray}
Then, let us interpret it in terms of differential forms on $M$.

The following correspondence is obvious one:
\begin{eqnarray}
  \{H,\quad\}_{1}=\xi_{H}^i \frac{\partial}{\partial \theta^i}
& \rightarrow &\hat\imath _{H} -{\rm operator \quad of\quad inner\quad
product} \nonumber\\ && {\rm  on\quad the\quad vector\quad
 field}\quad \xi_{H} ,\nonumber\\
\{F,\quad\}_{1}= \theta^i
 \frac{\partial}{\partial x^i}  & \rightarrow & \hat d -{\rm operator\quad
of\quad
 exterior\quad differentiation},\nonumber\\
\{ Q,\quad \}_{1}=\xi_{H}^i
 \frac{\partial}{\partial x^i} + \xi_{H,k}^{i}\theta^k
\frac{\partial}{\partial \theta^i} &\rightarrow & \hat{\cal L}_{H} -{\rm
Lie\quad derivative\quad along}\quad \xi_{H} .\nonumber\\
\label{eq:corr}\end{eqnarray}
 Using Jacobi identities (\ref{eq:bjac}) we
obtain
\begin{equation} \{H,F\}_1=Q_H\rightarrow \hat d\hat \imath _H+\hat
\imath _H\hat d=\hat{\cal L}_H-{\rm homotopy\quad formula}.\nonumber
\end{equation}

As we see, the supercharge $H+F,$ which defines the supersymmetry
transformation (\ref{eq:susy}), corresponds to the operator of equivariant
differentiation.

Now let us allow that some Riemannian metric $g_{ij}$ is defined on $M$ and
$\xi _H^i$ is its Killing vector.\\
Then the ''gauge fermion'' (\ref{eq:gauge}) is the motion
integral of the
odd mechanics $(Q_H, \Omega_1, {\cal M})$:
\begin{equation}
\label{eq:killing}{\cal L}_Hg=0-{\rm {Killing\quad equation}}\rightarrow
\{Q_H,{\tilde Q}_H\}_1=0.
\end{equation}

The functions $F$ and $H$ commute with ${\tilde Q}$ by the following way:
\begin{equation}
\label{eq:FQ,HQ}\{F,{\tilde Q}_H\}_1=-F_2,\quad \{H,Q_H\}_1=H_2\,\,\,\,\,,
\end{equation}
where
\begin{equation}
\label{eq:F1H1}H_2=\xi _H^ig_{ij}\xi _H^j,\quad F_2=\frac 12\theta ^i\omega
_{(2)ij}\theta ^j,\quad \omega _{(2)ij}=\frac{\partial (g_{ik}\xi _H^k)}{%
\partial x^j}-\frac{\partial (g_{jk}\xi _H^k)}{\partial x^i}.
\end{equation}

It is easy to check up that the mechanics $(H,\omega ,M)$ and $(H_2,\omega
_2 ,M)$ define the same Hamiltonian vector field on $M$ (it was
shown at first in Ref. \cite{niemi3}):
 \begin{equation}
\label{eq:bH}\frac{\partial H}{\partial x^i}\omega ^{ij}=\frac{\partial H_2}{%
\partial x^i}\omega _2^{ij} .
 \end{equation}

Obviously, one can separate two different cases:

 i) The case, when
these Hamiltonian systems coincides with each other (up to a constant
multiplier):
\begin{equation}
H=H_2,\quad \omega =\omega _2.
 \label{hh}\end{equation}
Then $H,F,Q_H,{\tilde Q}_H$ form the closed superalgebra
\begin{eqnarray}
 && \{H\pm F, H\pm F\}_{1} =\pm 2Q_{H} ,
 \{H\pm F,{\tilde Q}_{H} \}_{1}= H\mp F ,\\
 && \{H+ F, H -  F \}_{1} = \{H\pm F, Q_{H} \}_{1}
= \{Q_{H}, {\tilde Q}_{H} \}_{1} = 0  .                         \nonumber
\end{eqnarray}
It coincides with the superalgebra of 1D Witten's mechanics.

In the case of path integral generalization (see below) this corresponds to
the topological field theory \cite{niemi2}.

ii) These Hamiltonian systems are different ones:
\begin{equation}
H\neq H_2,\quad \omega \neq \omega _2.
\end{equation}
Then $(H_2,\omega _{2},M)$ defines the second Hamiltonian structure.

If the Poisson brackets corresponded to $\omega $ and $\omega _2$ satisfy
the compatibility condition (e. i. if any linear combination of that satisfy
the Jacobi identity) then the initial mechanics $(H,\omega ,M)$ is
integrable one. \\

Due to $M$ provided with both symplectic and Riemannian structures we can
define on the supermanifold ${\cal M}$ the {\it even symplectic structure}.
For this purpose let us consider on ${\cal M}$ the following local 1-form:
\begin{equation}
\label{eq:sA}{\cal A}_\alpha =A_{(\alpha )i}dx^i+\theta ^ig_{ij}D\theta ^j.
\end{equation}
Here $A_\alpha =A_{(\alpha )i}dx^i$ is local (pre)symplectic 1-form on $M$:
$$
dA_\alpha =\omega _\alpha ,
$$
and
$$
D\theta ^i=d\theta ^i+\Gamma _{kl}^i\theta ^kdx^l,
$$
where $\Gamma _{kl}^i$ are Cristoffel symbols for metrics $g_{ij}$ on $M$).

It is easy to see that the exterior differential of this 1-form is {\it %
globally } defined on ${\cal M}$ and represents  the following symplectic
structure:
\begin{equation}
\label{eq:Oalpha}{\tilde \Omega }_\alpha =d{\cal A}_\alpha =\frac 12(\omega
_{(\alpha )ij}+R_{ijkl}\theta ^k\theta ^l)dx^i\wedge dx^j+g_{ij}D\theta
^i\wedge D\theta ^j,
\end{equation}
where $R_{ijkl}$--curvature tensor on $M$ (of course, this is not the unique
form of an even symplectic structure on $%
{\cal M}$) .

The Poisson bracket corresponded  to this structure is the following one:
\begin{equation}
\label{epb}\{f(z),g(z)\}_\alpha =\nabla _if(z)(\omega _{(\alpha
)ij}+R_{ijkl}\theta ^k\theta ^l)^{-1}\nabla _jg(z)+\frac 12\frac{\partial
_rf(z)}{\partial \theta ^i}g^{ij}\frac{\partial _lg(z)}{\partial \theta ^j},
\end{equation}
where $g^{ik}g_{kj}=\delta _j^i$,
$$
\nabla _i=\frac \partial {\partial x^i}-\Gamma _{ij}^k\theta ^j\frac{%
\partial _l}{\partial \theta ^k}.
$$

Let us assume in (\ref{eq:sA})--(%
\ref{epb}) : $\alpha =0,2,\quad\omega _0\equiv\omega ,\quad H_0\equiv H $.

Then it is easy to see that $({\cal H}_\alpha ,\Omega _\alpha ,{\cal M})$
and $(Q_H,\Omega _1,{\cal M})$   are defining the same Hamiltonian
dynamics on ${\cal M}$:
\begin{equation}
\{z^A,{\cal H}_\alpha \}_\alpha =\{z^A,Q_H\}_1 ,
\quad {\rm where}\quad {\cal H}_\alpha =H_\alpha +F_2,\quad
\quad \alpha=0,2\,\,.
 \end{equation}
Thus, we have shown that on the DH-supermanifold there exists the exotic
structure of bi-Hamiltonian dynamics with even and odd symplectic
structures. Such dynamics has been considered
in Refs.\cite{volkov}-\cite{red}.

Without loss of generality,  we can choose in presented constructions the
metric $g_{ij}$ by such a way that $\omega ^{ij}g_{jk}=I_k^i$ defines an
almost complex structure ( $I_l^iI_k^l=-\delta _k^i$) which is
Lie-invariant with respect to $\xi _H^i$ . In the case, if $\hat I$ is the
complex structure, the (super)manifolds $M$ and ${\cal M}$ are K\"ahlerian
ones. Moreover, ${\cal M}$ is provided by both even and odd K\"ahlerian
structures. Bi-Hamiltonian dynamics with even and odd symplectic structures
and BV-structures on such supermanifolds have been considered in \cite{tmp},%
\cite{red}.

It is obvious that such dynamics includes  the integrable systems on the
orbits of co-adjoint representation of semisimple Lie groups (which are
K\"ahlerian manifolds).

\setcounter{equation}0

\section{Equivariant Localization}

Now we demonstrate the derivation of DH-formula (\ref{DH}) using the
constructions presented above.

We can present the integral (\ref{eq:int2}) in the form
\begin{equation}
\label{eq:int3}Z_\lambda =\frac 1{(\pi )^N}\int_{{\cal M}}\exp {\
(H-F-\lambda \{H+F,{\tilde Q\}_1})}d^{4N}z,
\end{equation}
where $F$ and ${\tilde Q}$ are defined by the expressions (\ref{eq:F0}) , (%
\ref{eq:gauge}) and $\{\quad ,\quad \}_1$ by (\ref{eq:bxt}).

The vector fields (\ref{eq:corr}) conserve the ``volume form'' $%
d^{4N}z=d^{2N}xd^{2N}\theta $ ( or, equivalently, $\Delta _0H=\Delta
_0F=\Delta _0Q_H=0$). From (\ref{eq:sualg2}), (\ref{eq:killing}) we deduce
$$
\{H+F,{\rm e}^{(H-F-\lambda \{H+F,{\tilde Q})\}_1}\}_1=0,\quad \{Q,{\rm e}%
^{(H-F-\lambda \{H+F,{\tilde Q}\}_1)}\}_1=0.
$$
Therefore, the integral (\ref{eq:int3}) is invariant under equivariant and
Lie transformations along $\xi _H^i$. Using the definition of operator $%
\Delta $ (\ref{eq:delta}) we can present this fact in the form:
\begin{equation}
\Delta _{H-F}(H+F)=\Delta _{H-F}Q=0.
\end{equation}
We have also
$$
\{Q,{\tilde Q}{\rm e}^{(H-F-\lambda \{H+F,{\tilde Q}\}_1)}\}_1=0.
$$
Using these expressions and the fact that the integral of an equivariantly
exact form vanishes on the compact manifold we show :
\begin{eqnarray}
\frac{dZ_{\lambda}}{d\lambda}
&=&-\frac{\lambda }{\pi^N}
\int_{\cal M}\{H+F, {\tilde Q}\}_{1}
{\rm e}^{(H -F-\lambda \{H+F, {\tilde Q}\}_{1})} d^{4N}z =  \nonumber \\
&=&-\frac{\lambda }{\pi^N}\int_{\cal M}\{H+F, {\tilde Q}
{\rm e}^{(H -F- \lambda \{H+F, {\tilde Q}\}_{1})}\}_{1} d^{4N}z +\nonumber\\
&+&\frac{\lambda}{\pi^N}\int_{\cal M}{\tilde Q}\{H+F,
{\rm e}^{(H -F-\lambda \{H+F, {\tilde Q}\}_{1})}\}_{1} d^{4N}z =0 .\nonumber
\label{eq:lambda}\end{eqnarray}
Thus,taking the limits $\lambda \to 0$, $\lambda \to \infty $ and due to%
$$
\delta (\xi _H^i)=\frac 1{\pi ^N}\lim _{\lambda \to \infty }\sqrt{\lambda
^{2N}detg_{ij}}{\rm e}^{-\lambda \xi _H^ig_{ij}\xi _H^j},\nonumber
$$
we obtain the DH-localization formula:
\begin{eqnarray}
&Z_{0}&=\frac{1}{(2\pi)^N}\int_{M}{\rm e}^{H}\sqrt{det\omega_{ij}}d^{2N}x=
\lim_{\alpha\to\infty}\frac{1}{\pi^N}\int_{{\cal M}}{\rm e}^{(H -F-
\lambda (H_{2}-F_{2}))}d^{4N}z = \nonumber\\
&&=\int_{M}{\rm e}^{H}\delta{(\xi_{H})}\sqrt{\det\omega_{ij}}\sqrt{\det
\frac{\partial \xi_{H}^{i}}{\partial x^j}}d^{2N}x .
\label{DH2} \end{eqnarray}

The path integral generalization of the presented constructions one can
accomplish by the lifting it on the loop space, similarly to \cite{niemi1}-%
\cite{niemi3} .For this, we have to consider the integral in (\ref{int}) and
(\ref{eq:int3}) as the path integral over loop space $L{\cal M}$ with the
boundary conditions on $z^A$: $z^A(0)=z^A(T)\,$and replace the Hamiltonian $H
$ by action $S$ and lift on the loop space the constructions obtained in
previous Sections:
\begin{eqnarray}
 H&\rightarrow& iS=i\int_{0}^{T}{A_{i}dx^{i} - Hdt}, {\rm where}\quad
dA=\omega ,
\nonumber\\
 F(z)&\rightarrow&
 F^{L}(z(t))=-\int_{0}^{T}dt\omega_{ij}(x(t))\theta^{i}(t)\theta^{j}(t)
,\nonumber\\
g_{ij}(x)dx^{i}dx^{j}& \rightarrow& \int_{0}^{T}dt
g_{ij}(x(t))dx^{i}(t) dx^{j}(t) ,\nonumber\\
\{f(z), g(z)\}_{1}&\rightarrow&
\{f(z(t)), g(z(t))\}^{L}_{1}=
\int_{0}^{T}dt\frac{\delta_{r} f(z(t))}{\delta z^{A}}
\Omega_{(1)AB}(z(t)) \frac{\delta_{l} g(z(t))}{\delta z^{A}},\nonumber
\end{eqnarray} where $\Omega _{(1)AB}(z(t))$ is defined by (\ref{eq:bxt}).

Then, we get
\begin{eqnarray}
\xi^{i}_{H}&\rightarrow&\xi^{i}_{S}=\{ x^{i}(t), S\}^{L}_{1}=
\dot{x}^i -\xi^{i}_{H},\nonumber\\
 Q&\rightarrow& Q_{S}=\{S, F^{L}\}^{L}_{1}=
\int_{0}^{T}dt\xi^{i}_{S}\omega_{ij}(x(t))\theta^{j}(t) ,\nonumber\\
{\tilde Q}&\rightarrow& {\tilde Q_{S}}=
\int_{0}^{T}dt \xi^{i}_{S}(x(t))g_{ij}(x(t))\theta^{i}(t),\nonumber
\end{eqnarray}
and the argument of $\delta $-function in the last integral in (\ref{DH2})
changes from $\xi _H^i$ to $\xi _S^i$, i. e. the path integral localizes to
the ordinary integral over the classical phase space.

\setcounter{equation}0

\section{Conclusion}

We have shown that the supermanifold of Duistermaat-Heckman localization can
be provided with :
\begin{enumerate}
\item  the structures of Batalin-Vilkovisky formalism, namely, with odd
symplectic structure and nilpotent operator $\Delta $;
\item  the structure of supersymmetric bi-Hamiltonian dynamics with even and
odd symplectic structures.
\end{enumerate}
These structures allow one to describe the equvariant localization by
Hamiltonian way without introducing the additional structure of cotangent
bundle of supermanifold ( associated with tangent bundle of the initial
manifold), and thus - without introducing the additional $4N$ variables.

Using the first structure one can consider the derivation of the degenerate
DH-formula (including path integral generalized case) via BV-formalism .

The second structure gives the example of dynamical application of the odd
symplectic structure. It establishes the correspondence between initial
dynamics and its supersymmetrization and gives also the supersymmetrization
way for the wide class of integrable systems. One can consider this
correspondence  in a quantum level too, taking into account that replacing
in path integral generalization of (\ref{eq:int3})
$$
\{S-F^L,{\tilde Q}_S\}_1^L\rightarrow \{H-F,{\tilde Q}_S\}_1^L
$$
one obtains the partition function of the supersymmetric dynamics,
considered above (the case (\ref{hh}), if $\lambda \neq 0,\infty $ .

Let us notice that the representation of the initial integral in the form (%
\ref{int}) formally coincides to the form of the integral from
differential forms on a {\it supermanifold} \cite{ber} and the presented
description is a symmetrical one according to the initial and auxiliary
coordinates. Thus, it can be generalized for the super-Hamiltonian systems.
\section{Acknowledgments}
I am very thankful to A.I.Batalin and T.Voronov for
useful discussions , and to E.A.Ivanov, S.O.Krivonos, M.A.Mukhtarov,
A.I.Pashnev for interest to work.

\end{document}